\documentclass[conference]{IEEEtran}
\IEEEoverridecommandlockouts
\usepackage{cite}
\usepackage{amsmath,amssymb,amsfonts}
\usepackage{algorithmic}
\usepackage{graphicx}
\usepackage{textcomp}
\usepackage{xcolor}
\usepackage[T1]{fontenc}
\def\BibTeX{{\rm B\kern-.05em{\sc i\kern-.025em b}\kern-.08em
    T\kern-.1667em\lower.7ex\hbox{E}\kern-.125emX}}
    
\usepackage{hyperref} 
\usepackage{setspace}

\makeatletter
\def\@IEEEpubidpullup{8\baselineskip}
\makeatother

\begin{document}

\IEEEoverridecommandlockouts
\IEEEpubid{
\parbox{\columnwidth}{\vspace{-4\baselineskip}Permission to make digital or hard copies of all or part of this work for personal or classroom use is granted without fee provided that copies are not made or distributed for profit or commercial advantage and that copies bear this notice and the full citation on the first page. Copyrights for components of this work owned by others than ACM must be honored. Abstracting with credit is permitted. To copy otherwise, or republish, to post on servers or to redistribute to lists, requires prior specific permission and/or a fee. Request permissions from \href{mailto:permissions@acm.org}{permissions@acm.org}.\hfill\vspace{-0.8\baselineskip}\\
\begin{spacing}{1.2}
\small\textit{ASONAM '21}, August 27-30, 2021, The Hague, The Netherlands \\ 
\copyright\space2021 Association for Computing Machinery. \\
ACM ISBN 978-1-4503-9128-3/21/11/\$15.00 \\
\url{http://dx.doi.org/10.1145/3487351.3488277} 
\end{spacing}
\hfill}
\hspace{0.9\columnsep}\makebox[\columnwidth]{\hfill}}
\IEEEpubidadjcol


\title{Identifying Influential Nodes Using Overlapping Modularity Vitality}


\author{\IEEEauthorblockN{Stephany Rajeh}
\IEEEauthorblockA{
\textit{University of Burgundy}\\
Dijon, France \\
Email: stephany.rajeh\\@u-bourgogne.fr}
\and
\IEEEauthorblockN{Marinette Savonnet}
\IEEEauthorblockA{
\textit{University of Burgundy}\\
Dijon, France \\
Email: marinette.savonnet\\@u-bourgogne.fr}
\and
\IEEEauthorblockN{Eric Leclercq}
\IEEEauthorblockA{
\textit{University of Burgundy}\\
Dijon, France \\
Email: eric.leclercq\\@u-bourgogne.fr}
\and
\IEEEauthorblockN{Hocine Cherifi}
\IEEEauthorblockA{
\textit{University of Burgundy}\\
Dijon, France \\
Email: hocine.cherifi\\@u-bourgogne.fr}
}

\maketitle

\begin{abstract}
It is of paramount importance to uncover influential nodes to control diffusion phenomena in a network. In recent works, there is a growing trend to investigate the role of the community structure to solve this issue. Up to now, the vast majority of the so-called community-aware centrality measures rely on non-overlapping community structure. However, in many real-world networks, such as social networks, the communities overlap. In other words, a node can belong to multiple communities. To overcome this drawback, we propose and investigate the “Overlapping Modularity Vitality” centrality measure. This extension of “Modularity Vitality” quantifies the community structure strength variation when removing a node. It allows identifying a node as a hub or a bridge based on its contribution to the overlapping modularity of a network. A comparative analysis with its non-overlapping version using the Susceptible-Infected-Recovered (SIR) epidemic diffusion model has been performed on a set of six real-world networks. Overall, Overlapping Modularity Vitality outperforms its alternative. These results illustrate the importance of incorporating knowledge about the overlapping community structure to identify influential nodes effectively. Moreover, one can use multiple ranking strategies as the two measures are signed. Results show that selecting the nodes with the top positive or the top absolute centrality values is more effective than choosing the ones with the maximum negative  values to spread the epidemic.

\end{abstract}


\section{Introduction}
Amidst the decisiveness of curbing the spread of epidemics, improving marketing and awareness campaigns, or maintaining network connectivity, identifying influential nodes in the dynamics of networks is a challenging issue. Centrality measures are one of the main approaches to rank nodes according to their importance in the network. Classical centrality measures rely on local or global information on the network topology \cite{lu2016vital}. For example, Degree centrality measures the number of links of a node. It relies only on the knowledge of the node's neighborhood. In contrast, Betweenness centrality uses information of the whole network. Indeed, it measures the frequency of a node lying in the shortest paths between all other nodes. Generally, local measures are computationally efficient but less accurate, while it is the opposite for global measures. These two types of measures are also combined to get the best of both worlds \cite{ibnoulouafi2018m,bucur2020top}.

The modular organization of real-world networks is one of the essential topological features. They are formed with modules of densely connected nodes that have rare connections with the other modules. Nonetheless, nodes can belong to several modules. Indeed, many networks, such as social, collaboration, biological, and infrastructural networks, are characterized by an overlapping community structure \cite{palla2005uncovering, xie2013overlapping, jebabli2014overlapping,jebabli2015user}. In these networks, the overlapping nodes may have multiple functions. For example, a person may adhere to several organizations. A scientist may conduct interdisciplinary research with various scientific groups. A protein may have several functions in different complexes. Also, an airport may have internal and external flights. Indeed, an overlapping community structure is frequent in many real-world situations.

Even though real-world networks are generally modular, classical centrality measures discard this important feature. Most of the centrality measures proposed in the literature are agnostic about the community structure, such as Degree centrality, Betweenness centrality, and $k$-core \cite{das2018study, bloch2019centrality, rajeh2020interplay, kitsak2010identification}. A new emergent research area exploits the community structure to identify influential nodes \cite{modvitality, ghalmane2019immunization, ghalmane2019centrality, tulu2018identifying, gupta2016centrality, guimera2005functional, zhao2015community, luo2016identifying}. They quantify the local and global influence of a node through its intra-community links and inter-community links \cite{rajeh2021characterizing}. The former connects nodes in the same community, while the latter connects nodes in different communities.
Nonetheless, these measures consider that a node belongs to a unique community. For example, Community-based Mediator \cite{tulu2018identifying} targets influential nodes based on the entropy of their intra-community links and inter-community links. There are no overlapping nodes. Indeed, if a node is overlapping, its centrality measure needs to be evaluated depending on its multiple communities. One can build various strategies accordingly.
Another example is Comm Centrality \cite{gupta2016centrality}. It targets hubs and bridges simultaneously but gives more weight to bridges. Here too, a node belongs to one and only one community. Few works have shown that incorporating information on the overlapping community structure results in identifying influential nodes more effectively \cite{ghalmane2019centralityoverlapping, chakraborty2016immunization, taghavian2017local}.

A recent work investigates the behavior of a set of non-overlapping community-aware centrality measures. Results show that Modularity Vitality \cite{modvitality} outperforms its alternatives. This signed community-aware centrality measure quantifies the modularity variation when one removes a node in the network. It can pinpoint bridges or hubs depending on the ranking scheme. Indeed, eliminating hubs in a community decreases modularity. In contrast, removing bridges between communities increases the modularity.
As it has proved to be quite effective, we propose to extend it to networks with an overlapping community structure. The SIR infectious spreading model is used to evaluate the performance of "Overlapping Modularity Vitality" compared to its non-overlapping version. Simulations are performed on six real-world networks originating from three domains: infrastructural networks, collaboration networks, and online social networks. Results demonstrate that Overlapping Modularity Vitality outperforms its non-overlapping counterpart. Moreover, when considering the absolute value of both measures (i.e., targeting hubs and bridges simultaneously), the difference in the performance of the two measures version is even higher.

The paper is organized as follows. The non-overlapping and overlapping versions of Modularity Vitality are given in section \ref{sec:MV}. The data and the tools used in the evaluation process are presented in section \ref{sec:DataAndTools}. In section \ref{sec:ExpResults}, experimental results are given. A discussion is developed in section \ref{sec:Disc}. Finally, section \ref{sec:Conc} concludes the article.

\section{Non-overlapping and Overlapping Modularity Vitality}
\label{sec:MV}
In this section, we present the definitions of non-overlapping Modularity Vitality based on Newman's modularity and its overlapping version. First, suppose that $G(V,E)$ is an undirected and unweighted graph where $V$ is the set of nodes of size $N=|V|$ and $|E|$ is the total number of edges. The connections between the nodes are described in the graph's adjacency matrix $A$. The graph $G$ is partitioned into $C=\{c_1, c_2, ... , c_k, ... c_{|C|}\}$ communities where $c_k$ is $k$-th community and $|C|$ is the total number of communities. In a network with a non-overlapping community structure, node $v_i$ can belong to one and only one community. In a network with an overlapping community structure, node $v_i$ can belong to several communities. That being said, its strength of belonging differs from one community to another. Hence, each node $v_i$ is characterized by a belonging coefficient vector $(\alpha_{v_i, c_1}, \alpha_{v_i, c_2}, ..., \alpha_{v_i, c_k}, ..., \alpha_{v_i, c_{|C|}})$ and assuming the following conditions hold: $0 \leq a_{v_i, c_k} \leq 1; \forall v_i \in V; \forall c_k \in C$ and $ \sum\limits_{c_k \in C} a_{v_i, c_k} = 1$.

\subsection{Non-overlapping Modularity Vitality}
Modularity Vitality \cite{modvitality} is based on Newman's modularity \cite{newman2006modularity}.  A lot of community detection algorithms use this property as an objective function \cite{blondel2008fast, clauset2004finding, yang2016modularity, shang2013community}. It measures the difference between the actual connections within a community to the connections with wiring at random. The larger the value of modularity, the stronger the community structure. Hence, developing a community-aware centrality measure that is directly related to the community structure strength is ideal. Newman's modularity is defined as follows:
\begin{equation}
Q(G) =  \sum_{c_k \in C}  \left[ \frac{|E_{c_k}^{in}|}{|E|} - \left(\frac{2|E_{c_k}^{in}| + |E_{c_k}^{out}|}{2|E|} \right)^2\right]
\end{equation}

where:
\begin{itemize}
  \item $|E_{c_k}^{in}| = \frac{1}{2} \sum\limits_{v_i, v_j \in c_k} A_{v_i,v_j}$, the intra-community edges of $c_k$
  \item $|E_c^{out}|= \sum\limits_{v_i  \in c_k} \sum\limits_{v_j  \in C - c_k} A_{v_i,v_j}$, the inter-community edges of $c_k$ 

\end{itemize}




Using the above definition of modularity, Modularity Vitality can distinguish hubs from bridges based on their contribution to the overall modularity of a network. Indeed, hubs increase the modularity while bridges decrease it. It is defined as follows:
\begin{equation}
\alpha_{MV}(v_i) = Q(G) -  Q(G\setminus\{v_i\})
\end{equation}
where: 
\begin{itemize}
  \item $Q(G)$ is the network's modularity based on Newman
\item $Q(G\setminus\{v_i\})$ is the network's modularity based on Newman after the removal of node $v_i$

\end{itemize}
Since Modularity Vitality is signed, we can use three ranking strategies. The first ranking strategy ranks nodes in decreasing order from the highest positive to the lowest negative value. This strategy targets hubs first. The second orders nodes from negative to positive centrality values. Consequently, it targets bridges first. Finally, nodes are ordered according to their absolute centrality value. It allows targeting hubs and bridges simultaneously based on their contribution to the network's overall modularity.

\subsection{Overlapping Modularity Vitality}
Several extensions to Newman's modularity for overlapping community structures have been proposed. One can divide them into two groups: fuzzy modularity and non-fuzzy modularity \cite{chen2015fuzzy, lazar2010modularity}. In the former, a node belongs to several communities based on a belonging coefficient. In the latter, the relationship is binary. This paper uses a simple fuzzy version of modularity to quantify the quality of the overlapping community structure. It is defined as follows:
\begin{equation}
a_{v_i,c_k} =  \frac{1}{O_{v_i}}
\end{equation}

where $O_{v_i}$ is the number of communities shared by node $v_i$. In other words, if a node belongs to one community, its belonging coefficient is 1. Node shared by two communities, gets a belonging coefficient of 0.5 for each of the communities, and so on.

Based on the belonging coefficient vector for each node $v_i$, this information can be incorporated within $|E_{c_k}^{in}|$ and $|E_{c_k}^{out}|$, where:
\begin{equation}
|E_{c_k}^{in}| = \frac{1}{2} \sum\limits_{v_i, v_j \in c_k} \frac{a_{v_i,c_k} + a_{v_j,c_k}}{2} A_{v_i,v_j}
\end{equation}

\begin{equation}
|E_{c_k}^{out}| = \sum\limits_{v_i  \in c_k} \sum\limits_{v_j  \in C-c_k} \frac{a_{v_i,c_k}+(1-a_{v_j,c_k})}{2}A_{v_i,v_j}
\end{equation}
  
Hence, the main difference between the overlapping version and the non-overlapping version is that the adjacency matrix ($A_{v_i,v_j}$) contains the weights of the average of belonging coefficients of the nodes. Indeed, it has been shown that the definition of modularity holds for a non-binary adjacency matrix \cite{newman2004analysis}. The Overlapping Modularity Vitality is defined as follows:
\begin{equation}
\alpha_{OMV}(v_i) =   Q_o(G) - Q_o(G\setminus\{v_i\})
\end{equation}
where:
\begin{itemize}
  \item $Q_o(G)$ is the network's overlapping modularity
\item $Q_o(G\setminus\{v_i\})$ is the network's overlapping modularity after the removal of node $v_i$
\end{itemize}

Similar to its non-overlapping version, Overlapping Modularity Vitality is also signed. Consequently, we also consider three ranking strategies as defined for its non-overlapping counterpart.

\section{Data and Tools}
\label{sec:DataAndTools}

This section presents the real-world networks used, the community detection algorithms, as well as the SIR evaluation process, and the subsequent evaluation criterion to compare the community-aware centrality measures.

\subsection{Data}
We perform a comparative evaluation on six real-world networks originating from three domains: infrastructural, collaboration, and online social networks. Table \ref{TableBasicTopology} reports their basic topological characteristics.

\subsubsection{Infrastructural Networks}
We consider two air transportation networks, EU Airlines \cite{netz}, and U.S. Airports \cite{kunegis2014handbook}. In these networks, the nodes represent European and U.S. airports, respectively. The nodes are connected if there's a direct flight between them.

\subsubsection{Collaboration Networks} 
The AstroPh \cite{nr} network is extracted from the e-print arXiv. Nodes represent authors who have submitted their papers to the Astrophysics category. The nodes are connected if two people have co-authored a paper. In New Zealand Collaboration \cite{netz}, the nodes are scientific institutions in New Zealand such as universities and organizations. If Scopus lists a minimum of one publication with authors in any two institutions, the nodes are connected.

\subsubsection{Online Social Networks}
Hamsterster \cite{kunegis2014handbook} is an online social pet platform. Nodes represent users of the platform, and connections represent online friendships. In the DNC Emails \cite{netz}, nodes represent members of the Democrat National Committee. If the members have exchanged emails, they are connected.

\begin{table*}[h]
\caption{Basic topological properties of the networks. $N$ is the number of nodes. \textit{|E|} is the number of edges. $<k>$ is the average degree. $\zeta$ is the transitivity. $Q$ is the non-overlapping modularity. $Q_o$ is the overlapping modularity. $on (\%)$ is the fraction of overlapping nodes. $m$ is the average number of community memberships of the nodes. The overlapping properties are based on the community structure uncovered by SLPA. * means the largest connected component of the network is taken if it is disconnected.}
\begin{center}
\begin{tabular}{|p{2.1cm}|c|c|c|c|c|c|c|c|}
\hline
\textbf{Network} & $N$ & $|E|$ & $<k>$ & $\zeta$ & $Q$ & $Q_o$ & $on (\%)$ & $m$\\
\hline
EU Airlines & 417 & 2,953 &  14.16 & 0.304 & 0.109 & 0.741  & 0.062 & 2.154\\
U.S. Airports & 500 & 2,980 & 11.92 & 0.351 & 0.161  & 0.731  & 0.118 & 2.152 \\
DNC Emails* & 849 & 10,384 & 24.46 & 0.548 & 0.416 & 0.593  & 0.285 & 2.004 \\
New Zealand* & 1,463 & 4,246 & 5.80 & 0.063 & 0.401 & 0.524 & 0.364 & 2.163\\ 
Hamsterster* & 1,788  & 12,476 & 13.49 & 0.090 & 0.391 & 0.648   & 0.251 & 2.247 \\
AstroPh* & 17,903 & 196,972 & 22.00 & 0.317 & 0.563 & 0.208 & 0.569 & 2.669\\

\hline

\end{tabular}
\label{TableBasicTopology}
\end{center}
\end{table*}

\subsection{Community Detection Algorithms}
As the network's community structure is unknown, one uses a community detection algorithm to uncover it. Infomap \cite{rosvall2008maps} reveals non-overlapping communities while Speaker-Listener Label Propagation Algorithm (SLPA) \cite{xie2011slpa} unveils overlapping communities. These influential algorithms have proved their effectiveness in estimating the mesoscopic scale \cite{orman2012comparative, jebabli2018community, dao2018community}.

\subsubsection{Infomap} extracts a non-overlapping community structure by minimizing the description of a random walk. Densely connected modules that are sparsely connected characterize many real-world networks. Therefore, it is likely for a random walker to stay longer within the modules than jumping from one module to another. One uses Huffman coding with a prefix code and a suffix code. The prefix code refers to the module. The suffix code is assigned to a node within the modules. The compression of the description of the random walker unveils the community structure.

\subsubsection{SLPA} extracts overlapping communities. It is inspired by how humans acquire and spread opinions. As a start, each node has a unique label. Then, each node acts as a listener and selects one of the labels sent to it from its neighbors (i.e., speakers). The listener updates its memory based on the label frequency it listens to. When all the nodes are visited, the label frequency in their memory is converted to a probability distribution. This distribution characterizes the membership degree for each of the communities. The distribution of the membership of each node is further processed to extract the communities. A threshold $r$ allows discarding the membership to a specific community if it is below the threshold. In the experiments, we set its value to 0.01.


\subsection{Susceptible-Infected-Removed Model}
\label{sec:SIR}
One uses the epidemiological model Susceptible-Infected-Recovered (SIR) \cite{anderson1979population} to assess the effectiveness of the community-aware centrality measures. In this model, nodes can be in either of these states: Susceptible (S), Infected (I), or Recovered (R). Initially, all nodes are susceptible except for a given proportion ($f_o$) of the top nodes ranked according to a specific centrality measure. These nodes are infected. Each infected node can infect its susceptible neighbor with a probability $\lambda$. Simultaneously, each infected node can recover with a probability $\gamma$. The propagation proceeds until all nodes are either recovered or still in the susceptible state. At this point, one computes the outbreak size, which is the number of nodes in the recovered state (R). This outbreak size dictates the spreading effectiveness of a specific centrality measure for each $f_o$ considered. Centrality measures aim to maximize this value. In the experiments, 100 SIR simulations are performed and averaged for each fraction of initially infected nodes ($f_o$) for the community-aware centrality measures under study.

\subsection{Evaluation Measure}
We use the Degree centrality as a baseline to compare the spreading outbreak size of community-aware centrality measures. The relative difference is defined as:
\begin{equation}
\Delta R = \frac{R_c - R_b}{R_b}
\end{equation}


where:
\begin{itemize}
  \item $R_c$ denotes the outbreak size using a specific community-aware centrality measure $c$
\item $R_b$ represents the outbreak size using the baseline Degree centrality
\end{itemize}

If the community-aware centrality measure is more effective than the baseline, $\Delta R$ is positive. Otherwise, $\Delta R$ is negative.

\section{Experimental Results}
\label{sec:ExpResults}
This section reports the comparative evaluation of the effectiveness of the non-overlapping Modularity Vitality and its overlapping version.

\subsection{Targeting Hubs}
First, we examine how Modularity Vitality in its non-overlapping and overlapping versions compare when the fraction of initially infected nodes ($f_o$) is chosen based on the ranks ordered from the highest positive centrality values. In other words, initially infected nodes are local hubs as they contribute positively to modularity.  The left column of figure \ref{Fig-SIR} shows the relative difference of the outbreak size ($\Delta R$) as a function of the fraction of initially infected nodes ($f_o$) where the nodes are ranked based on the top positives (hubs-first). One observes three typical behaviors.

The first case is illustrated by EU Airlines, U.S. Airports, Hamsterster, and AstroPh. In these networks, Overlapping Modularity Vitality ($\alpha_{OMV}$) outperforms its non-overlapping version ($\alpha_{MV}$). However, the outperformance of $\alpha_{OMV}$ starts after infecting 9\% of the nodes in EU Airlines and AstroPh and after 5\% in U.S. Airports and Hamsterster. This behavior demonstrates that infecting local hubs is more beneficial than infecting hubs located in the overlap when resources are limited. Above this fraction, the outperformance of $\alpha_{OMV}$ compared to $\alpha_{MV}$ can reach up to 5\%. In this situation, hubs in the overlap matter. The second typical behavior is illustrated by DNC Emails. In this network, the curves of the relative difference of the outbreak size ($\Delta R$) for both $\alpha_{OMV}$ and $\alpha_{MV}$ essentially intersect when $f_o >$ 0.10. In other words, $\alpha_{OMV}$ and $\alpha_{MV}$ perform similarly. Finally, the third behavior is illustrated by the New Zealand Collaboration network. In this network, the non-overlapping Modularity Vitality ($\alpha_{MV}$) outperforms its overlapping version ($\alpha_{OMV}$) when $f_o >$ 0.26. Below this value, the curves mainly coincide.

\subsection{Targeting Bridges}
\begin{figure*}[ht!]
\centerline{\includegraphics[width=8 in, height=8.5 in]{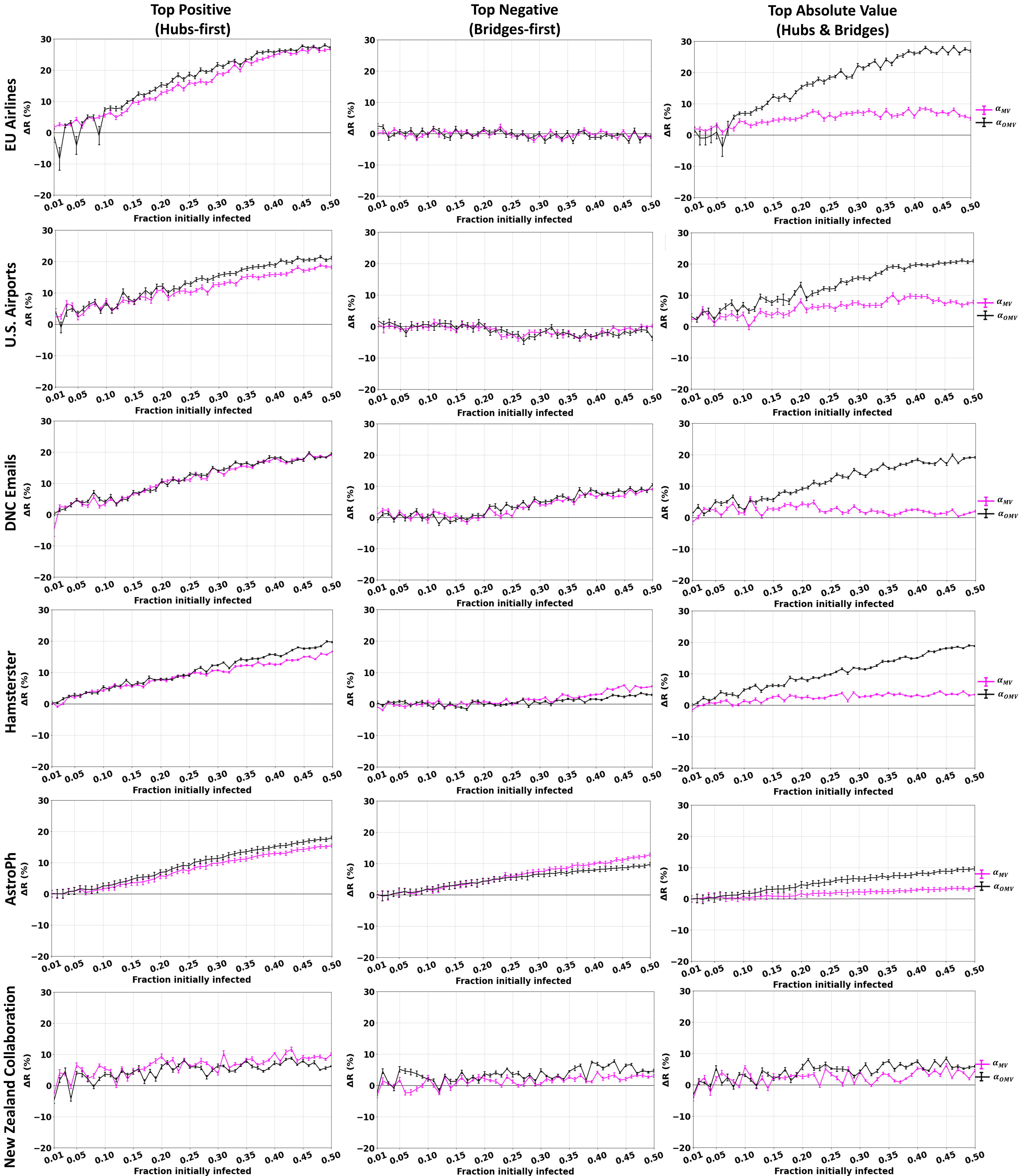}}
\caption{Relative difference of the outbreak size ($\Delta R$) as a function of the fraction of initially infected nodes ($f_o$) on the six real-world networks used. The community-aware centrality measures are Modularity Vitality ($\alpha_{MV}$) and Overlapping Modularity Vitality ($\alpha_{OMV}$). On the left, middle, and right of the figure, nodes are ordered from top positive centrality values, top negative centrality values, and top absolute values, respectively.}
\label{Fig-SIR}
\end{figure*}

The performances of the non-overlapping Modularity Vitality ($\alpha_{MV}$) and its overlapping version ($\alpha_{OMV}$) are investigated on the SIR model when the fraction of initially infected nodes ($f_o$) contains the nodes ordered from the highest negative centrality values. In other words, the initially infected nodes are bridges since they contribute negatively to modularity. The middle column of figure \ref{Fig-SIR} shows the relative difference of the outbreak size ($\Delta R$) as a function of the fraction of initially infected nodes ($f_o$) where the nodes are ranked based on the top negatives (bridges-first). Once again, one can distinguish three behaviors.

EU Airlines and U.S. Airports illustrate the first behavior. In these networks, the curves of the relative difference of the outbreak size ($\Delta R$) for both $\alpha_{OMV}$ and $\alpha_{MV}$ essentially intersect at all fractions of initially infected nodes. Moreover, they don't exhibit a higher performance compared to the baseline. They even perform worse than the baseline in the U.S. Airports network after infecting 20\% of the nodes. Indeed, $\Delta R$ is negative in this case. Hamsterster and AstroPh illustrate the second behavior. In these two networks, $\alpha_{MV}$ results in a higher epidemic outbreak than its overlapping version when $f_o >$ 0.27 for Hamsterster and when $f_o >$ 0.24 for AstroPh. Its outperformance can reach up to 4\% when compared to $\alpha_{OMV}$. When $f_o$ is less than these two values, the curves generally coincide. Finally, the third behavior is illustrated by the networks DNC Emails and New Zealand Collaboration. Here, the Overlapping Modularity Vitality shows a slightly better performance than its non-overlapping version. However, similar to Hamsterster and AstroPh, when $f_o$ is less than 0.16 in DNC Emails, both measures share a similar relative difference of the epidemic outbreak size.

\subsection{Targeting Hubs and Bridges}
In the final experiment, we investigate the performances of the non-overlapping Modularity Vitality ($\alpha_{MV}$) and its overlapping version ($\alpha_{OMV}$) when the fraction of initially infected nodes ($f_o$) uses the absolute value of the centrality measures ranks. In other words, the initially infected nodes target hubs and bridges simultaneously, based on their global effect on modularity. The right column of figure \ref{Fig-SIR} shows the relative difference of the outbreak size ($\Delta R$) as a function of the fraction of initially infected nodes ($f_o$) where the nodes are ranked based on the absolute value (hubs and bridges).

Infecting nodes based on their global effect, regardless of their nature (hubs or bridges), the Overlapping Modularity Vitality ($\alpha_{OMV}$) outperforms its non-overlapping counterpart in all of the networks under study. Indeed, the outperformance of $\alpha_{OMV}$ compared to $\alpha_{MV}$ can reach up to 28\% in terms of the relative difference of the outbreak size ($\Delta R$) in EU Airlines. Up to 19\% in DNC Emails and Hamsterster. Up to 13\% in U.S. Airports. Up to 7\% in AstroPh. Finally, up to 5\% in New Zealand Collaboration. The network where $\alpha_{MV}$ and $\alpha_{OMV}$ are the closest in terms of their relative difference is New Zealand Collaboration. In the other networks, the difference between $\alpha_{MV}$ and $\alpha_{OMV}$ increases as $f_o$ increases. Indeed, one can see that when the fraction of initially infected nodes is low, the difference between $\alpha_{OMV}$ and $\alpha_{MV}$ is minimal. This illustrates the effectiveness of incorporating information from the overlapping community structure and not discarding the influence of hubs on behalf of bridges (and vice versa). In other words, both hubs and bridges play a crucial role in epidemic spreading. Targeting both of them leads to a higher diffusion.

\section{Discussion}
\label{sec:Disc}
In this work, we introduce an extension of Modularity Vitality ($\alpha_{MV}$), a community-aware centrality measure designed for networks with non-overlapping community structure to overlapping communities. The Overlapping Modularity Vitality ($\alpha_{OMV}$) addresses a critical gap of many realistic situations where a node might belong to several communities. The extension we consider uses a simple weighting scheme based on the reciprocal number of communities shared by a node. This information is incorporated in the adjacency matrix of the network. One computes the weight by averaging the strength of membership to different communities of the overlapping nodes. This strategy is inspired by one of the various strategies discussed in \cite{chen2015fuzzy}. So doing, hubs shared by several communities contribute the most to the overlapping modularity of a network. Conversely, bridge nodes belonging to several communities have a substantial negative contribution to the network's overlapping modularity.

Since both community-aware centrality measures can target bridges and hubs, we investigate the effectiveness of three different ranking strategies in a SIR epidemic scenario. 
First, nodes are ranked in decreasing order of their centrality. This strategy targets hubs at first. With this ranking scheme, when the fraction of initially infected nodes ($f_o$) is less than 9\%, there is no difference between the overlapping and the non-overlapping centrality measures ($\alpha_{MV}$ and  $\alpha_{OMV}$). These results corroborate similar findings reported in \cite{ghalmane2020exploring}. Indeed, extensive analysis shows that most hubs are direct neighbors of overlapping nodes. Hence, the epidemic outbreak effect is not tremendously different since local hubs and overlapping hubs are in the vicinity of each other. The difference is more pronounced when we infect more than 9\% of the nodes. Here, $\alpha_{OMV}$ targets overlapping hub nodes that are more strategically positioned in a network, resulting in a higher epidemic outbreak.

Second, we investigate the strategy consisting of infecting nodes with the highest negative contribution to the non-overlapping ($Q$) and overlapping modularity ($Q_o$). In other words, we now target bridges. We notice that non-overlapping Modularity Vitality ($\alpha_{MV}$) and Overlapping Modularity Vitality ($\alpha_{OMV}$) are less effective. Indeed, even a negative relative difference of the outbreak size ($\Delta R$) is observed. In addition, the highest relative difference of the outbreak size amounts to 13\% when compared to the baseline for $\alpha_{MV}$ and 10\% for $\alpha_{OMV}$. It occurs in the AstroPh network. These values are almost half of the ones observed when the ranking scheme targets hubs first. Indeed, when hubs are infected first, the highest relative difference of the outbreak size reaches 27\% when compared to the baseline for the Overlapping Modularity Vitality and 26\% for the non-overlapping Modularity Vitality. Hence, targeting bridges is not as effective as targeting hubs. Moreover, one can see in DNC Emails and New Zealand Collaboration networks that $\alpha_{OMV}$ has a slightly higher performance than $\alpha_{MV}$. For Hamsterster and AstroPh, it is the opposite. Yet in both cases, when the fraction of initially infected nodes ($f_o$) is less than 0.27, it is hard to differentiate between $\alpha_{MV}$ and $\alpha_{OMV}$. These results suggest that bridges identified by the two measures are very similar or close to each other. Consequently, they result in identical relative differences in the outbreak size.

Finally, we investigate infecting the nodes based on the absolute value of the centrality values. This strategy simultaneously targets hubs and bridges. In this case, the Overlapping Modularity Vitality ($\alpha_{OMV}$) outperforms its counterpart on all networks under study. Moreover, the relative difference of the epidemic outbreak size reaches up to 28\% in EU Airlines. This is the highest value reached among all networks throughout the three different ranking schemes. The outperformance of $\alpha_{OMV}$ demonstrates the importance of considering the roles of overlapping hubs and overlapping bridges simultaneously rather than targeting each in silo.

It is worth mentioning deviations from the general trend. In some networks, using non-overlapping Modularity Vitality ($\alpha_{MV}$) leads to a higher epidemic spreading compared to its overlapping version ($\alpha_{OMV}$). It happens when targeting hubs first in the New Zealand Collaboration network and when targeting bridges first in Hamsterster and AstroPh. These networks are characterized with a high proportion of overlapping nodes ($on (\%)$ = 0.364, 0.251, 0.561, respectively). In these cases, $\alpha_{OMV}$ might not be able to differentiate between strategically positioned hubs and bridges. However, when targeting both (i.e., taking the absolute value), $\alpha_{OMV}$ shows its merit. It indicates the importance of targeting hubs and bridges simultaneously rather than opting for one or the other. However, these preliminary results need more investigation on networks with different proportions of overlapping nodes.

\section{Conclusion}
\label{sec:Conc}
Many real-world networks naturally have a community structure made of communities tightly connected internally and loosely connected externally. In a lot of realistic situations, nodes can belong to several communities. These nodes, called overlapping nodes, frequently occur in social networks, collaboration networks, biological networks, ecological networks, to name a few. For example, one protein may contribute to various functions, one human may belong to multiple institutions, and one researcher may work in several scientific domains.

Identifying influential nodes within these networks is a crucial issue. Indeed, in networks with a community structure, nodes can act as hubs (locally influential) or bridges (globally influential). This problem gave rise to several propositions in the literature. However, Modularity Vitality is the only one directly linked to the quality of the community structure. It can target hubs and bridges by quantifying the node's contribution to the modularity. In this work, we propose an extension of Modularity Vitality incorporating information about the overlap in communities. Using SIR simulations, we show that Overlapping Modularity Vitality outperforms its counterpart designed for non-overlapping communities. The difference between both measures is higher when one uses the absolute value of the centrality to rank the nodes. These results demonstrate the importance of incorporating information about the overlapping communities to identify influential nodes better. Moreover, results show that it is more effective to target hubs or hubs and bridges simultaneously rather than targeting bridges.

In future work, we plan to investigate the various alternative definitions of overlapping modularity. As in many situations, the network's community structure is unknown; one needs to evaluate the robustness of the results to community structure variation linked to the community detection algorithm. Furthermore, we plan to relate the overall performance of the Overlapping Modularity Vitality with the network topological properties.

\bibliographystyle{IEEEtran}
\bibliography{IEEEabrv,bibtech}

\end{document}